\documentclass[%
 reprint,
 amsmath,amssymb,
 aip,twocolumn
]{revtex4}

\usepackage{graphicx}
\usepackage{dcolumn}
\usepackage{bm}
\usepackage[frozencache=true,cachedir=minted-cache]{minted}

\usepackage[utf8]{inputenc}
\usepackage[T1]{fontenc}
\usepackage{mathptmx}
\usepackage{etoolbox}
\usepackage{tcolorbox}
\usepackage{color,soul}
\usepackage{listings}
\makeatletter
\def\@email#1#2{%
 \endgroup
 \patchcmd{\titleblock@produce}
  {\frontmatter@RRAPformat}
  {\frontmatter@RRAPformat{\produce@RRAP{*#1\href{mailto:#2}{#2}}}\frontmatter@RRAPformat}
  {}{}
}%
\makeatother

\usepackage[utf8]{inputenc}
\usepackage{amsmath}
\usepackage{url}
\usepackage[colorlinks=true]{hyperref}

\usepackage{hyperref}
\usepackage{algorithm,algpseudocode}
\usepackage{pgf}
\usepackage{pgfplots}
\pgfplotsset{compat=1.17}
\usepackage{tikz}
\usepackage{soul} 
\usepackage{amsthm,amsfonts}
\usepackage{mathtools} 
\usepackage{fancyvrb} 
\usepackage{media9}
\usepackage{epigraph}
\usepackage{smartdiagram}
\usepackage{verbatim}
\usepackage{lmodern}
\usepackage{listings}
\usepackage{tcolorbox}
\usepackage{color,soul}
\usepackage{minted}
\usepackage{natbib}
\usepackage{graphicx}
\usepackage{listings}

\newcommand{\vv}{\mathbf{v}}

\begin{document}

\title{Hybrid programming-model strategies for GPU offloading of electronic structure calculation kernels}

\author{Jean-Luc Fattebert}
\email{fattebertj@ornl.gov}
\affiliation{
Computational Sciences and Engineering Division,  Oak Ridge National Laboratory
}

\author{Christian F. A. Negre}
\affiliation{Theoretical Division, Los Alamos National Laboratory}

\author{Joshua Finkelstein}
\affiliation{Theoretical Division, Los Alamos National Laboratory}
 
\author{Jamaludin Mohd-Yusof}
\affiliation{Computer, Computational, and Statistical Sciences Division, Los Alamos National Laboratory}

\author{Daniel Osei-Kuffuor}
\affiliation{Center for Applied Scientific Computing, Lawrence Livermore National Laboratory}

\author{Michael E. Wall}
\affiliation{Computer, Computational, and Statistical Sciences Division, Los Alamos National Laboratory}
 
\author{Yu Zhang}
\affiliation{Theoretical Division, Los Alamos National Laboratory}

\author{Nicolas Bock}
\affiliation{Canonical Ltd.}

\author{Susan M. Mniszewski}
\affiliation{Computer, Computational, and Statistical Sciences Division, Los Alamos National Laboratory}

\date{\today}

\begin{abstract}
To address the challenge of performance portability, and facilitate the implementation of electronic structure solvers, we developed the Basic Matrix Library (BML) and Parallel, Rapid O(N) and Graph-based Recursive Electronic Structure Solver (PROGRESS) libraries. BML implements linear algebra operations necessary for electronic structure kernels using a unified user interface for various matrix formats (dense, sparse) and architectures (CPUs, GPUs). Focusing on Density Functional Theory (DFT) and Tight-Binding (TB) models, PROGRESS implements several solvers for computing the single-particle density matrix and relies on BML. In this paper, we describe the general strategies used for these implementations on various computer architectures, using OpenMP target functionalities on GPUs, in conjunction with third-party libraries to handle performance critical numerical kernels. We  demonstrate the portability of this approach and its performance on benchmark problems.
\end{abstract}

\maketitle

\section{Introduction} 


Performance portability is a significant challenge for application programs that are run on modern HPC resources. For example, software solutions targeting portability such as OpenMP sometimes can have a hard time delivering performance, either due to the lack of maturity of compilers or the fine granularity of control needed for some specific kernels. On the other hand, writing kernels in a vendor specific language targeting one specific GPU may not be portable and can lead to software maintenance difficulties. Recent trends in HPC only make this problem more acute, as many leadership computing facilities have adopted hardware composed of heterogeneous compute nodes containing CPUs and GPUs, where a majority of the acceleration is provided by the GPUs \cite{Deakin2022,Summit,Frontier}. 

In practice, it is usually best to use existing libraries when possible if those libraries implement the numerical kernels one needs. For electronic structure applications, linear algebra libraries are the most common dependency. For dense linear algebra on CPUs, standard interfaces developed for BLAS\cite{blas1,blas2,blas3} and Lapack\cite{lapack} have facilitated the use and development of these libraries, and several well optimized solutions exist. On GPUs, the situation is more complicated. Vendors offer optimized implementations of BLAS and Lapack in platform specific libraries such as cuBLAS, cuSolver (Nvidia), rocBLAS, rocSolver (AMD) and MKL (Intel). There is however no common interface to these libraries. As a consequence, application codes need to have platform specific wrappers around the functions they intend to use. In addition, application developers need to understand the details of all these interfaces. There are several reasons why the GPU situation is not as user friendly as it is on CPUs. First, there is the choice of having two possible locations for data arrays passed as arguments, either allocated on the host or the device, as well as for the return value, when there is one. Then there is also the option of enabling several kernels to execute asynchronously on the device, using for instance GPU streams.
When writing platform specific GPU kernels, there also can\ be extra kernel arguments depending on the architecture to ensure optimality such as, for instance, different hardware/run time parameters, different thread-block grid sizes or user controlled cache (shared memory) sizes. As a result of these various issues and GPU technology changes, user interfaces are not as stable as one would like. 
For sparse linear algebra, the situation is also complicated. Unlike the dense format for which there are not too many ways of laying out the data, there are several sparse formats including compressed sparse row (CSR), compressed sparse column (CSC), coordinate list (COO), and ELLPACK just to mention a few. In addition, even for a given format such as CSR, there are variants, as some libraries may or may not expect data in a row to be ordered by column indexes. 

Some open source alternatives have emerged in recent years, some of which are targeting multiple architectures and thus facilitating portability of application codes using those.
At the node level, the MAGMA library \cite{magma2014,magma2020} offers the functionalities of BLAS and Lapack on GPUs, and is already fully functional on Nvidia and AMD GPUs.
The SLATE project is implementing a distributed and scalable dense linear algebra library for distributed memory accelerator-based computer systems, aiming to provide performance and portability to various hardware
(CPUs, GPUs, accelerators) \cite{slate2022}.
On the sparse matrix side, Ginkgo\cite{ginkgo-toms-2022} is being actively developed and already offers a lot of functionalities on GPU architectures.

With the development of the Parallel, Rapid O(N) and Graph-based Recursive Electronic Structure Solver (PROGRESS) and the Basic Matrix Library (BML), our goal is to facilitate the development of performant and portable electronic structure solvers by providing the necessary linear algebra tools in a hardware agnostic way.
By providing several matrix formats, specifically a dense format and several sparse formats, BML facilitates the development of reduced complexity algorithms that can exploit any possible sparsity of the Hamiltonian and density matrix.
We reported on this concept and the BML library a few years ago in Ref.~\onlinecite{Bock2018}, with a focus on CPU implementations. In this paper we extend this concept to implementations on GPUs, demonstrating some of these ideas on various hardware such as Nvidia V100, AMD MI250X and Intel GPUs.
We describe in particular the implementation model used to offload calculations to GPUs, using OpenMP in combination with third-party libraries.
Electronic structure calculations are an important class of applications that require heavy use of linear algebra kernels. Here, electronic structure calculations refers broadly to the many ways of numerically evaluating the state of electrons in a physical system (molecule, periodic solid, etc.) as necessary to derive other physical quantities of interest.
In this paper we will restrict our discussion to mean-field models such as Density Functional Theory (DFT) or Tight-Binding methods.
Some algorithms and implementations discussed are targeting large scale simulations and make use of matrix sparsity to reduce computational complexity to $\mathcal{O}(N)$.
Moreover, fast time-to-solution is also of high interest in the community, specifically to speedup wall-clock times in quantum molecular dynamics (QMD) and enable better modeling with longer time-scales for medium-size systems, on the order of 1,000 electrons.

The idea of isolating all the linear algebra operations of an electronic structure code into a separate library is a natural design choice, and at the same time allows for multiple application codes to share this implementation. Several other research groups have made efforts towards identifying and isolating software libraries and have made them available to the community. The DBCSR library\cite{dbcsr,Schutt2016}, which the CP2K simulation package\cite{cp2k} relies on, is designed to efficiently perform sparse matrix-matrix multiplication, among other operations. It provides a distributed implementation using MPI and runs on Nvidia and AMD GPUs via CUDA and HIP, respectively. The ELectronic Structure Infrastructure (ELSI) project \cite{elsi} provides an open-source software interface to facilitate the implementation and optimal use in electronic structure codes of high-performance solver libraries including traditional eigensolvers, $\mathcal{O}(N)$ complexity algorithms, and other reduced complexity methods.
The Electronic Structure Library (ESL) \cite{esl2020} is a community-maintained library of software specific to electronic structure simulations, which includes among others the ELSI library just mentioned.
On the application side, an example of an electronic structure code that recently embraced this separation of operations and the use of more libraries is SIESTA \cite{Lebedeva2023}.


After introducing the PROGRESS and BML libraries in Section \ref{se:bml}, we discuss in Section \ref{se:dm} the specific problem of electronic structure these libraries are targeting: computing the single-particle density matrix (DM).
We then discuss in Section \ref{se:gpu} some GPU-friendly algorithms implemented in PROGRESS as possible alternatives to a direct dense diagonalization.
In Section \ref{se:openmp}, we describe our general strategy to offload computational kernels to the GPU using OpenMP.
We then describe some more specific strategies used for the dense matrix format in Section \ref{se:dense} (using the MAGMA and MKL libraries) and sparse matrix format in Section \ref{se:sparse} (using AMD rocSparse library and the Hypre library).
Finally in Section \ref{se:dist}, we discuss distributed memory solvers that leverage the shared memory solvers discussed in previous sections.

\section{PROGRESS and BML libraries}
\label{se:bml}

The Basic Matric Library (BML) is designed to implement the linear algebra operations necessary to implement matrix-based electronic solvers.
Its purpose is to hide all the implementation details of numerically intensive kernels, including architecture specific code and interfaces with third-party libraries, from the user.

BML is written in C. This facilitates interoperability with other languages such as Fortran and C++.
BML supports four datatypes for all operations: single precision, double precision, single complex, double complex.
To avoid writing essentially the same code for all the data types, C macros are used extensively for data types and in function names, and the C preprocessor is used to generate specifics code associated with each data type. 
To avoid naming conflicts with other packages, all BML function names use a prefix $\verb|bml_|$.

BML supports four different matrix formats: dense, ELLPACK-R, CSR and ELLBLOCK.
ELLPACK-R is a sparse format with a fixed memory allocation allowing a pre-determined maximum number $M$ of non-zero elements per row \cite{ellpack-r}.
It is less adaptable than CSR since a growing number of non-zeroes may reach the limit $M$ and lead to a failure. But it has the performance advantage of avoiding a lot of memory allocation/deallocation during routine usage.
ELLBLOCK is a generalization of ELLPACK-R where each element is a block\cite{CoPA2021}.
Note that BML CSR format implementation is not exactly what is usually referred to as CSR. In BML, rows of data and column indexes are stored independently of each other instead of being packed together into a single array. This facilitates memory reallocation when the number of non-zeroes in a row changes.  
In addition, BML supports a memory distributed format, ``distributed2d'' which builds on top of the four non-distributed formats.

BML matrices are stored as a C struct, which includes, besides a pointer to the data storage, other parameters necessary to fully describe that matrix, such as its type (e.g. dense or ELLPACK-R), its data type (float, double, complex), and the number of rows in the matrix. An example for an ELLPACK-R matrix is shown in Listing \ref{list:matrix}.

\begin{listing}[!ht]
\begin{minted}{c}
struct bml_matrix_ellpack_t
{
    /* matrix type identifier. */
    bml_matrix_type_t matrix_type;
    /* data type */
    bml_matrix_precision_t matrix_precision;
    /* number of rows. */
    int N;
    /* number of columns in storage */
    int M;
    /* matrix elements */
    void *value;
    /* column indexes */
    int *index;
    /* number of non zeros per row */
    int *nnz;
    ...
}
\end{minted}
\caption{C struct used for ELLPACK-R matrix storage}
\label{list:matrix}
\end{listing}

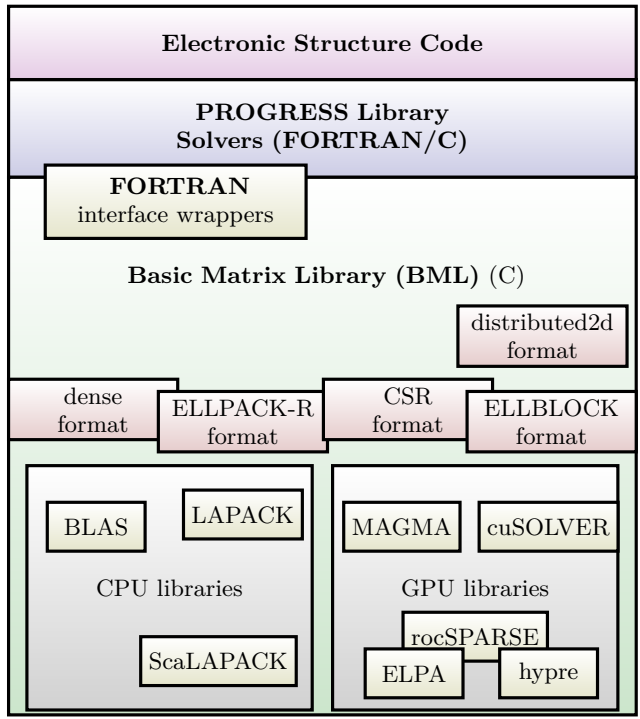
\begin{figure}[ht]
     \centering
     \begin{tikzpicture}[->,shorten >=1pt,auto,
         block_app/.style ={rectangle, draw=black, 
             very thick, fill=white,
             draw=black,
             top color=white,
             bottom color=magenta!50!black!20,
             text width=25em, text centered,
             minimum height=3em},
         block_prg/.style ={rectangle, draw=black, 
             very thick, fill=white,
             draw=black,
             top color=white,
             bottom color=blue!50!black!20,
             text width=25em, text centered,
             minimum height=4em},
         block_bml/.style ={rectangle, draw=black, 
             very thick, fill=white,
             draw=black,
             top color=white,
             bottom color=green!50!black!20,
             text width=25em, text centered,
             minimum height=22em},
         block_lib/.style ={rectangle, draw=black, 
             very thick, fill=white,
             draw=black,
             top color=white,
             bottom color=white!10!black!20,
             text width=11em, text centered,
             minimum height=10em},        
         block_cpu/.style ={rectangle, draw=black, 
             very thick, fill=white,
             draw=black,
             top color=white,
             bottom color=yellow!50!black!20, 
             text width=10em, text centered,
             minimum height=3em},        
         block_gpu/.style ={rectangle, draw=black, 
             very thick, fill=white,
             draw=black,
             top color=white,
             bottom color=red!50!black!20, 
             text width=12em, text centered,
             minimum height=4em},        
         block_formats/.style ={rectangle, draw=black, 
             very thick, fill=white,
             draw=black,
             top color=white,
             bottom color=red!50!black!20, 
             text width=6.25em, text centered,
             minimum height=2em},  
         block_tpl/.style ={rectangle, draw=black, 
             very thick, fill=white,
             draw=black,
             top color=white,
             bottom color=yellow!50!black!20, 
             minimum width=4em, text centered,
             minimum height=2em},           main node/.style={circle,draw,font=\Large\bfseries},
         node distance=0em,
         thick,        
         ]
         \node[block_prg] (prg) {\textbf{PROGRESS Library \\ Solvers (FORTRAN/C)}};
         \node[block_bml] (bml) [below of=prg,yshift=-13em] {\textbf{\vspace{12em}  Basic Matrix Library (BML)} (C)};
         \node[block_cpu] (wra) [below of=prg,yshift=-3em,xshift=-6em] {\vspace{0em} \textbf{FORTRAN} interface wrappers};   
         \node[block_lib] (cpulib) [below of=bml,yshift=-5.8em,xshift=-6.25em] {CPU libraries};
         \node[block_lib] (gpulib) [below of=bml,yshift=-5.8em,xshift=6.25em] {GPU libraries};
         \node[block_app] (app) [above of=prg,yshift=3.5em,xshift=0em] {\textbf{Electronic Structure Code}};
         \node[block_formats]  [below of=bml,yshift=4.5em,xshift=9em] {distributed2d \\ format};
         \node[block_formats]  [below of=bml,yshift=1.5em,xshift=-9.375em] {dense \\ format};
         \node[block_formats] (mkl) [below of=bml,yshift=1em,xshift=-3.25em] {ELLPACK-R \\ format};
         \node[block_formats]  [below of=bml,yshift=1.5em,xshift=3.5em] {CSR \\ format};
         \node[block_formats]  [below of=bml,yshift=1em,xshift=9.375em] {ELLBLOCK \\ format};
        \node[block_tpl][below of=cpulib,yshift=2.5em,xshift=-3em]{BLAS};
        \node[block_tpl][below of=cpulib,yshift=3em,xshift=3em]{LAPACK};
        \node[block_tpl][below of=cpulib,yshift=-3em,xshift=2em]{ScaLAPACK};
        \node[block_tpl][below of=gpulib,yshift=2.5em,xshift=-3em]{MAGMA};
        \node[block_tpl][below of=gpulib,yshift=2.5em,xshift=3em]{cuSOLVER};
        \node[block_tpl][below of=gpulib,yshift=-2.em]{rocSPARSE};         
        \node[block_tpl][below of=gpulib,yshift=-3.5em,xshift=3em]{hypre}; 
        \node[block_tpl][below of=gpulib,yshift=-3.5em,xshift=-2.5em]{ELPA}; 
\end{tikzpicture}
     \caption{Software stack showing the PROGRESS and BML libraries and their integration within an electronic structure application.}
     \label{fig:stack}
 \end{figure}

The PROGRESS (Parallel, Rapid O(N) and Graph-based Recursive Electronic Structure Solver) library is a collection of algorithms used in electronic structure calculations, with a focus on iterative solvers based on matrix polynomials.
It is written mostly in FORTRAN, but also offers a C-interface for routines expected to be called directly by an application code.
More specifically, it implements several versions of the ``second-order spectral projection'' (SP2) DM solver\cite{sp2}, a Chebyshev polynomial expansion of the DM\cite{Goedecker1995}, as well as some iterative methods to compute the inverse square root of an overlap matrix as is often necessary in a non-orthogonal basis set or a non-orthogonal tight-binding model. All these implementations rely on BML matrices and their functionalities. They are mostly matrix format agnostic, and available for all data types available in BML.
Several of these algorithms show an $\mathcal{O}(N)$ computational complexity with the matrix size $N$ when sparse matrix formats are used, and an appropriate threshold is used to discard small matrix elements. Fig.~\ref{fig:stack} shows the software stack for a typical application using PROGRESS and BML, including third-party dependencies. The PROGRESS and BML library are both open source, licensed under the BSD 3-clause license, and available on GitHub\cite{2023bml,2023progress}.

 {}

\begin{figure}[ht]
\centering
\includegraphics[width=8cm]{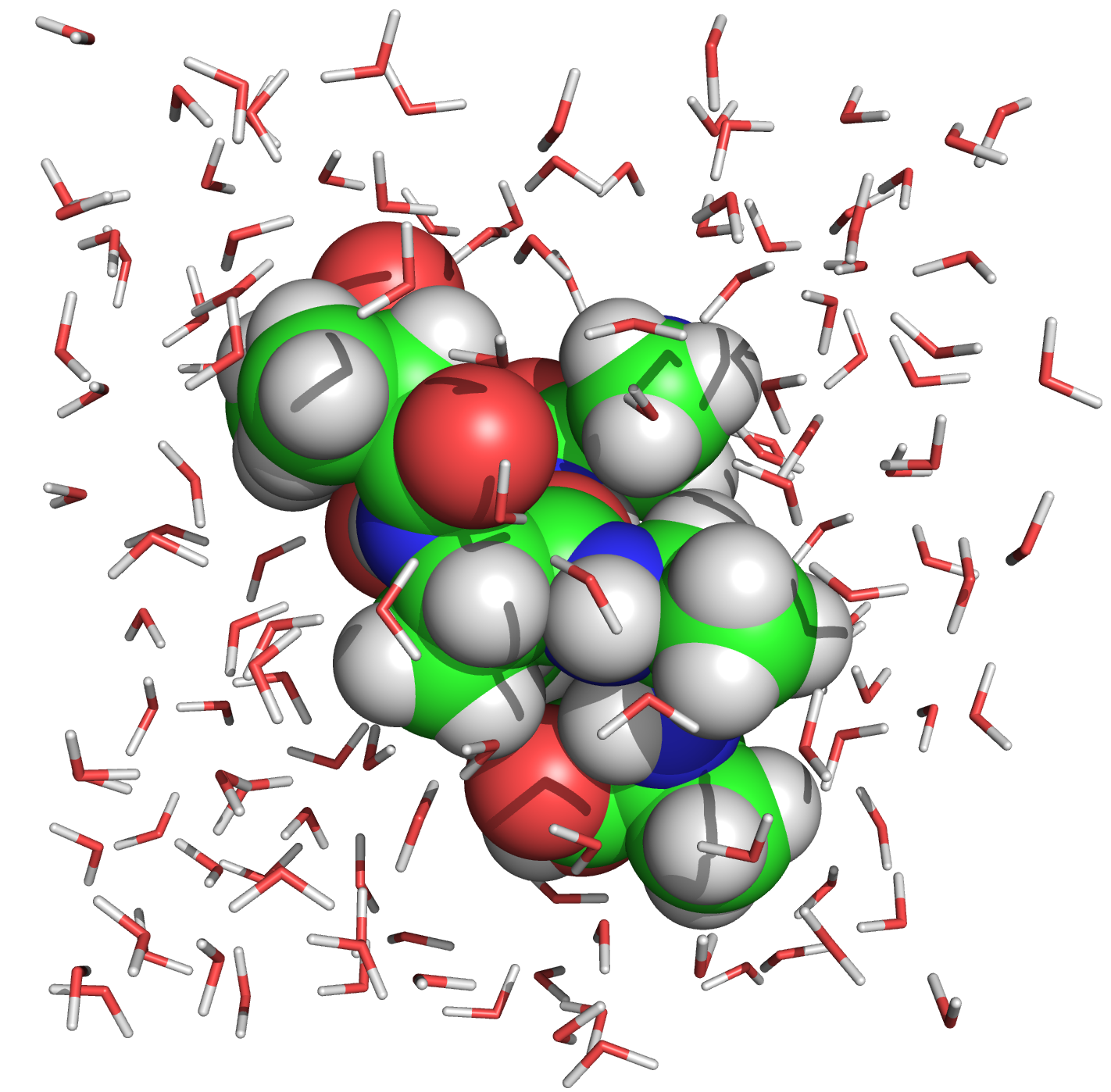}
\caption{Small peptide chain solvated in water, used for PROGRESS benchmarks. The figure was created using PyMOL\cite{pymol}.}
\label{fig:prot}
\end{figure}

To evaluate a solver's performance, we developed some benchmark drivers within PROGRESS.
Our main benchmark test is based on a physical system, a small peptide chain solvated in water with periodic boundary conditions (Fig.~\ref{fig:prot}).
It consists of 523 atoms.
From this atomic configuration, we build a tight-binding Hamiltonian represented by a matrix of size 1081$\times$1081.
To construct this Hamiltonian we call a Density Functional Based Tight Binding (DFTB)\cite{Frauhenheim2000} Hamiltonian builder implemented in PROGRESS that uses DFTB parameters from the LATTE (Los Alamos Transferable Tight-binding for Energetics) code\cite{latte} . 
We build larger Hamiltonians by replicating this system by a factor of two or three in each direction.
This gives us a series of Hamiltonians of 
increasing size to study computational cost, computational complexity, and parallel scaling. 
The sparsity of each DM for these benchmark problems are shown in Tab.~\ref{tab:sparsity}.
This resulting ``soft matter'' system is what one would typically encounter in a biophysical molecular dynamics (MD) simulation.

\begin{table}[ht!]
\begin{tabular}{|c|c| c|}
\hline
 Replicas & N & DM sparsity \\ 
 \hline
 1$\times$1$\times$1 & 1,081 & 57.5\% \\  
 2$\times$1$\times$1 & 2,162 & 78.7\% \\
 3$\times$1$\times$1 & 3,243 & 85.8\% \\
 2$\times$2$\times$1 & 4,324 & 89.3\% \\
 3$\times$2$\times$1 & 6,486 & 92.9\% \\
 2$\times$2$\times$2 & 8,648 & 94.6\% \\
 3$\times$3$\times$1 & 9,729 & 95.3\% \\
 3$\times$2$\times$2 & 12,972 & 96.4\% \\
 3$\times$3$\times$2 & 19,458 & 97.6\% \\
 3$\times$3$\times$3 & 29,187 & 98.4\% \\
 \hline
\end{tabular}
\caption{Percentage of zero-valued elements in DM in PROGRESS benchmark problem, using a cutoff threshold of $10^{-6}$.}
\label{tab:sparsity}
\end{table}

PROGRESS and BML rely on the CMake build system\cite{cmake}.
To facilitate the development of these libraries, and avoid breaking the code when changes are made by developers not familiar with all the functionalities and implementation details, an extensive suite of unit tests has been developed over time and continue to be enhanced. These tests cover all the matrix formats and matrix datatypes, and are run through Ctest, the testing driver provided by CMake.


\section{Single-particle density matrix and the associated eigenvalue problem}
\label{se:dm}

Given a symmetric (or Hermitian) $N\times N$ matrix $H$ representing a Hamiltonian operator in a finite basis set, the task of computing the single-particle density matrix $D$ in that same basis set can be accomplished by following a straightforward procedure

\begin{enumerate}
    \item Compute all the eigenvalues $\epsilon_i$ and eigenvectors $\vv_i$ of $H$.
    \item Given a chemical potential $\mu$, the Fermi-Dirac distribution function is given by
    \begin{equation}
    \label{eq:fd}
        f_\mu(\varepsilon) = \frac{1}{1 + \exp(\beta( \varepsilon-\mu))}.
    \end{equation}

    \item The single particle density matrix is given by
    \begin{equation}
    D=V 
    \begin{pmatrix}
    f_\mu(\varepsilon_1) &&& \\
     & f_\mu(\varepsilon_2) && \\
     && \dots & \\
     &&& f_\mu(\varepsilon_N)
    \end{pmatrix} V^T\;,
    \end{equation}
    where $V$ is the $N\times N$ matrix made of the ordered and orthonormal eigenvectors
    $$V=\begin{pmatrix}\vv_1 \vv_2 \dots \bf \vv_N\end{pmatrix}\;.$$
    
\end{enumerate}

Note that if there is a gap in the eigenvalue spectrum, and no eigenvalue close to $\mu$, $f_\mu(\varepsilon)$ takes either the value 1 ($\varepsilon < \mu$) or 0 ($\varepsilon > \mu$), provided $\beta$ is not too small. In this case we have an insulator. The value $\mu$ in practice can be determined by setting an occupation (number of electrons) and finding the $\mu$ which results in this occupation through an iterative process.
    
In practice, the Hamiltonian matrix can come, for example, directly from a discretization of the problem in a small basis set (such as a set of Gaussian-shaped orbitals centered on the atoms), through a parameterized tight-binding (TB) approximation, or from the projection of the Hamiltonian operator onto an auxilliary set of wave functions built iteratively during the search for the numerical solution in a larger {\it numerical} basis set (e.g. plane waves, finite elements or finite difference cases). Depending on the discretization of the problem and the solver adopted, the size and degree of sparsity of this matrix $H$ can vary significantly.

The matrix size also depends on the algorithm used to solve the electronic structure problem. When working with wave-functions, $N$ can be substantially smaller than in a TB or Gaussian-based approach. It can be as small as the number of occupied orbitals, as in a Conjugate Gradient solver\cite{CG89}, in which case no diagonalization is needed for the projected Hamiltonian (all states are fully occupied).
But wavefunction-based solvers often use $N$ larger than the number of occupied orbitals in order to speedup the solver, or to solve for partial occupancy in metallic systems\cite{KRESSE1996}.
In addition, solvers such as Block-Davidson will involve solving a Rayleigh-Ritz problem, often referred to in the field of electronic structure as subspace diagonalization, for $2N\times 2N$ matrices \cite{TACKETT2001,YANG2006,Fattebert_2022}. A Locally Optimal Block Preconditioned Conjugate Gradient (LOBPCG) solver\cite{BOTTIN2008} will involve an even larger space, with a Rayleigh-Ritz procedure in $3N\times 3N$ matrix space. 

Note that for wavefunction-based approaches, distribution of the wavefunction over nodes and cores can substantially reduce time-to-solution. However, the Rayleigh-Ritz process used to compute the orbitals occupation does involve information from all the distributed parts, so that the resulting synchronization communication step is often the bottleneck in the strong scaling limit \cite{Levitt2015,LUPOPASINI2020}, and thus requires efficient algorithms to solve that problem.

\section{GPU friendly algorithms} 
\label{se:gpu}

Matrix multiplications have two advantageous properties when it comes to their implementations on GPUs: (i) very simple arithmetic operations, (ii) high arithmetic intensity (floating point operations per memory load). When compared to the operations involved in solving a dense eigenvalue problem on a GPU, the use of standard dense diagonalization algorithms are typically not very efficient. Thus solvers based on matrix-matrix multiplications are able to better utilize the massively parallel threads on a GPU and may offer better performance \cite{CoPA2021,finkelstein2023fast,SKhadatkar23}.

Significant development in matrix-matrix multiplication based iterative solvers for the density matrix happened in the 1990's. Their primary purpose was to reduce algorithmic complexity from $\mathcal{O}(N^3)$ to $\mathcal{O}(N)$ by utilizing the sparsity in the Hamiltonian matrix and in functions of the Hamiltonian matrix  (see Ref.~\onlinecite{Goedecker1999} for a review). The key idea is to replace the diagonalization of the Hamiltonian matrix to evaluate the Fermi-Dirac function (see previous Section) with a much cheaper polynomial approximation that one can easily apply to a Hamiltonian matrix.
Often in these approaches, we define a shifted and rescaled Hamiltonian matrix 
$\widetilde{H}$
such that the eigenvalues of this matrix are all inside the interval $[a,b]$. A good polynomial approximation would then map the interval $[a,b]$ to a Fermi-Dirac function with the appropriate chemical potential $\mu$ so that
\begin{equation}
    D \approx p_{\mu}(\widetilde{H}) \;,
\end{equation}
where the subscript of the polynomial $p$ denotes the dependence on the chemical potential $\mu$. Another type of approximate expansion is a recursive polynomial expansion
\begin{equation}
    D \approx p_n(p_{n-1}(\dots p_2(p_1(\widetilde{H})))\;,
\end{equation}
so that the polynomial $p_\mu$ is replaced by a composition of generally simpler or lower-order polynomials: $p_\mu = p_n \circ p_{n-1} \circ \cdots \circ p_2 \circ p_1$. 

Although electronic structure problems can be quite large, domain scientists are often limited to solving problems of more modest sizes for which a very fast time-to-solution can be achieved. This is usually the case of quantum molecular dynamics where an electronic structure problem needs to be solved at each timestep to accurately compute atomic forces and propagate the atoms along the MD trajectories. In these more moderately-sized problem cases, even matrix multiplications cannot always fully utilize the available resources on GPU devices and substantial portions of the GPU remain idle. Therefore finding further parallelism in the evaluation of these polynomials is beneficial (see Section \ref{se:cheby}).

\subsection{SP2 solver}
\label{se:sp2}
An example of a recursive expansion is the ``second-order spectral projection'' (SP2) algorithm\cite{sp2} as implemented in PROGRESS. 
In SP2, one starts with 
\begin{equation}
D_0\equiv\widetilde{H}=\frac{\varepsilon_N I-H}{\varepsilon_N-\varepsilon_1} \;,
\end{equation}
the shifted and rescaled Hamiltonian, after which, the density matrix is computed iteratively using the recursion\begin{equation}
    D_{m+1}=D_m^2
\end{equation}
if the trace of $D_{m}$ is larger than the number of electrons, and
\begin{equation}
    D_{m+1}=2D_m-D_m^2
\end{equation}
if the trace of $D_{m}$ is smaller than the number of electrons. The density matrix $D$ is then approximated by $D_n$ for a sufficient number of iterations, $n$.
Listing \ref{list:sp2} shows a sketch of a Fortran code implementing an SP2 solver based on BML matrices and functionalities.
We should emphasize that such an implementation is independent of the matrix format, matrix data type, and underlying computer architecture.
Other variants of the SP2 algorithm have also been implemented in PROGRESS.

\begin{listing}[ht!]
\begin{minted}{fortran}
!SP2 solver to compute DM d_bml for Hamiltonian h_bml
!using a cutoff threshold tau
subroutine prg_sp2(h_bml,d_bml,tau,...)
  use bml
  !Declare BML matrices ...
  type(bml_matrix_t),intent(in) :: h_bml
  type(bml_matrix_t),intent(inout) :: d_bml
  type(bml_matrix_t) :: x2_bml
  ...
  !initialize d_bml as shifted, rescaled Hamiltonian
  call bml_copy(h_bml, d_bml)
  call bml_add_identity(d_bml,-1.*emax)
  call bml_scale(-1.0/(emax-emin), d_bml)
  !create BML matrix x2 as a copy of d_bml
  call bml_copy_new(d_bml, x2_bml)
  ...
  !SP2 loop
  do i=0,maxiter
    tr = bml_trace(d_bml)
    ! X2 <- X * X
    call bml_multiply_x2(d_bml,x2_bml,tau)
    if(tr-nel <= 0.) then
      call bml_add(2.,d_bml,-1.,x2_bml,tau)
    else
      call bml_copy(x2_bml, d_bml)
    end if
    !check for convergence
    ...
  end do
  call bml_deallocate(x2_bml)
end subroutine
\end{minted}
\caption{Illustration of use case for BML matrices in SP2 solver as implemented in PROGRESS library.}
\label{list:sp2}
\end{listing}

We use SP2 in our PROGRESS benchmark, specifically to evaluate performance of matrix-multiplication-based algorithms compared to dense diagonalization.
For the bio-system benchmark described in Section \ref{se:bml}, SP2 converges to the specific tolerance used in that benchmark in 22 iterations, and its cost is typically dominated by the 22 matrix-matrix multiplications used in these iterations.

\subsection{Chebyshev polynomial expansion}
\label{se:cheby}

Recursive Fermi operator expansion techniques, like SP2, have known challenges in the case of a vanishing electronic bandgap. A better technique in this case is to use a serial Chebyshev expansion of the Fermi operator\cite{Goedecker1995}
\begin{align}\label{eq:cheby_approx}
    f_\mu(H) \approx \sum_{n = 0}^L c_n T_n(H)\;,
\end{align}
where $T_n$ is the $n$th Chebyshev polynomial and $c_n$ is the $n$th Chebyshev expansion coefficient. Currently in PROGRESS, and in some other codes \cite{Mohr2017}, this expansion is computed via the Chebyshev polynomial recursion property
\begin{align}\label{eq:serial_cheby_recursion}
    T_{n+1}(H) = 2HT_n(H) - T_{n-1}(H)\;, \\ \nonumber
\end{align}
for $n>0$, which results in a serialized summation -- each term in the summation for $f_\mu$ needs to be computed in sequence. Moreover, each recursion requires a single matrix multiplication, so that for a Chebyshev expansion of polynomial order $L$, approximately $L$ matrix multiplications are required. In other words, the number of matrix multiplications to approximate the Fermi operator and compute the density matrix scales linearly with the number of expansion terms.

In 1973, Paterson and Stockmeyer \cite{Paterson1973} showed that a generic polynomial $P$ in powers of $x$ of order $L$ could be rewritten in such a way that only $\sim 2\sqrt{L}$ multiplications in $x$ are needed to evaluate $P(x)$. Several decades later, Liang et al. \cite{Liang2003,Liang2004} showed that this same idea could be applied directly to Chebyshev polynomial expansions of the Fermi-Dirac operator. The Chebyshev expansion in Eq.~(\ref{eq:cheby_approx}) can then be written as

\begin{align}
\begin{split}\label{eq:cheby_ansatz}
    \sum_{n=0}^L &c_n T_n  = \sum_{i=0}^{k-1} d_{i,0} T_i + T_k 
                \bigg(\sum_{i=0}^{k-1} d_{i,1} T_i \\ &+ T_k \bigg( \sum_{i=0}^{k-1} d_{i,2} T_i + \cdots + T_k \bigg( \sum_{i=0}^{k-1} d_{i,{m-1}} T_i \bigg) \cdots \bigg)\bigg)\;, 
\end{split}
\end{align}

for coefficients $d_{i,j}$, that can be determined from the Chebyshev coefficients $c_n$, and positive integers $k,m$ such that $L+1 = k m$. Not only did this algorithm replace a diagonalization with matrix-matrix multiplications, it also substantially reduced the \emph{number} of multiplications needed to approximate the Fermi-Dirac operator for a given Chebyshev expansion size. Instead of the nearly $L$ multiplications required for the serial recursion from Eq.~(\ref{eq:serial_cheby_recursion}), the number of multiplications now scaled with the square root of the polynomial order, $\sqrt{L}$. The large timing discrepancies between matrix-matrix multiplications and diagonalization previously mentioned then have an even more pronounced effect, making this approach very well-suited to modern GPU devices. The determination of the coefficients $d_{i,j}$ relies on the multiplicative recursion property of Chebyshev polynomials\cite{finkelstein2023fast}: for $n,m$ non-negative integers,  
\begin{align}\label{eq:mult_cheby_recursion}
    T_n T_m (H) = \tfrac{1}{2}(T_{n+m} + T_{|n-m|})(H)\;.
\end{align}

\begin{figure}
    \centering
    \begin{tikzpicture}
        \begin{semilogyaxis}[width=.485\textwidth,height=2.5in,
                    xmin=5, xmax=1050,
                    ymin=0.9, ymax=1000,
                    legend pos=north east,
                    legend columns=2,
                    ylabel near ticks,
                    ylabel={Speed up factor},
                    xlabel={Number of Chebyshev expansion terms},
                    ymajorgrids=true,
                    grid style=dashed,
                    domain=100:1000,
                    legend entries={$N=500$, 
                    $N=1000$,
                    $N=2000$, 
                    $N=4000$}]
        \addplot[blue,mark=square*,line width=0.5] table {data/crusher/500-full_speedup.txt};
        \addplot[brown,mark=*, line width=0.75] table 
        {data/crusher/1000-full_speedup.txt};
        \addplot[red,mark=triangle*, line width=1.0] table {data/crusher/2000x2000_diag_vs_PS-crusher.dat};
        \addplot[black,mark=diamond*, line width=0.75] table {data/crusher/4000x4000_diag_vs_PS-crusher.dat};
        \end{semilogyaxis}
    \end{tikzpicture}
    \caption{Speed up over diagonalization of the density matrix construction (using a known chemical potential) on an AMD MI250X GPU with mutliple GPU streams for $N=500$ and $N=1000$. Only a single stream is used for $N=2000$, $4000$. The MAGMA library is used for diagonalization which was found to be faster than the diagonalization routine in rocSolver.} \label{fig:cheby}
\end{figure}
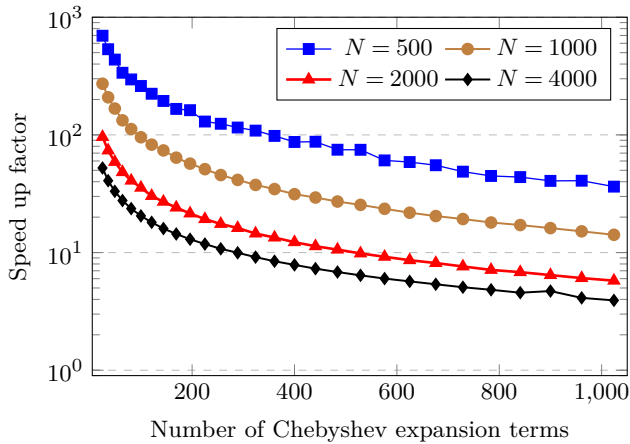

Note that each sum, $S_j$, on the right hand side of Eq.~(\ref{eq:cheby_ansatz}), where
\begin{align}\label{eq:cheby_sum}
    S_j = \sum_{i=0}^{k-1} d_{i,j} T_i\;, \qquad 0\le j \le m-1\;,
\end{align}
is completely independent of every other sum $S_{j'}$ for $j \ne j'$ once the first $k$ Chebyshev polynomials $T_i, i=0,\dots,k$ are known. This introduces parallelism into Eq.~(\ref{eq:cheby_ansatz}) since each sum can be calculated concurrently with every other sum. In addition, the first $k$ Chebyshev polynomials, $T_1$, $T_2$ up to $T_k$, can be calculated in a semi-parallel way too. The idea is to use the Chebyshev polynomial multiplication identity from Eq.~(\ref{eq:mult_cheby_recursion}) to calculate each $T_i$, $i=2,...,k$. Starting from $T_0$ and $T_1$, the second Chebyshev polynomial can be computed as
\begin{align}\label{eq:cheby1}
    T_2 = 2T_1T_1 - T_0\;,
\end{align}
and similarly, once $T_0$, $T_1$ and $T_2$ are known, the next two Chebyshev polynomials can be computed via:
\begin{align}
\begin{split}\label{eq:cheby2}
    T_3 & = 2T_2T_1 - T_1\\
    T_4 & = 2T_2T_2 - T_0\;,
\end{split}
\end{align}
using only previously computed $T_i$. Each Chebyshev polynomial on the left in Eq.~(\ref{eq:cheby2}) can therefore be computed in parallel. This parallelization on a single GPU is implemented using CUDA and HIP streams, on Nvidia and AMD GPUs, respectively, through MAGMA queues. The combination of this parallelization along with the square root scaling Chebyshev expansion approach leads to orders-of-magnitude speed ups over diagonalization when constructing the single-particle density matrix through a Chebyshev expansion. In Figure~\ref{fig:cheby}, we show the speed up obtained on a AMD MI250X GPU. 

Further details of this approach, and on how to compute the set of coefficients $\{d_{i,j}\}$, can be found in Ref.~\onlinecite{finkelstein2023fast}. 


\section{Offloading with OpenMP} 
\label{se:openmp}

There are two components of GPU offloading, data movement and compute. 
Since most accelerated architectures have separate memory spaces for host and device, with limited bandwidth between them, we attempt to minimize data movement whenever possible. 
OpenMP \cite{openmp,desupinski2018} offers a portable, pragma-based, framework for data movement between host and device, as well as compute functionality on both. The OpenMP standard is supported by multiple compiler vendors \cite{openmp_compilers}, with varying degrees of compatibility, especially with regard to offload support. 
GPU-offloading capabilities of OpenMP have been successfully used by various applications\cite{Bak2022,Thermo4PFM,SABAU2023}, including some in the electronic structure community\cite{qmcpack2022,gamess2023,gamess2023b}
and for large-scale sparse eigensolvers \cite{lobpcg2020}. 
In this section we outline the general strategy adopted in BML for using OpenMP to offload compute to accelerated devices. 

\subsection{Data Allocation}

The BML matrix format includes both the base matrix data (one or more data arrays, depending on the format) as well as certain metadata (e.g. matrix format, distribution mode, local array bounds) in a C struct (see Listing \ref{list:matrix}).
We only offload the data arrays to the device, at allocation time, using persistent data allocation on the device. For example, in the ELLPACK-R format, we offload the values, index and number of non-zeros arrays using a combination of allocation and updates from host to device (see Listing \ref{list:allocate_offload}). Subsequently, we assume that the data on the device is correct, and synchronization between device and host is only performed when needed.  We also modify the corresponding deallocation functions to ensure that device-side memory is freed when host-side arrays are destroyed. 

\begin{listing}[!ht]
\begin{minted}{c}
// allocate arrays on GPU
double* A_value=A->value;
#pragma omp target enter data map(alloc:A_value[:N*M])
int* A_index=A->index;
#pragma omp target enter data map(alloc:A_index[:N*M])
int* A_nnz=A->nnz;
#pragma omp target enter data map(alloc:A_nnz[:N])

// copy data from CPU to GPU
#pragma omp target update to(A_value[:N*M])
#pragma omp target update to(A_index[:N*M])
#pragma omp target update to(A_nnz[:N])
\end{minted}
\caption{GPU memory allocation and update for BML matrix data}
\label{list:allocate_offload}
\end{listing}

\subsection{OpenMP Compute}

For simple computations, or those computations which are not performance-critical, we can use OpenMP to perform device computations. For example, scaling an ELLPACK-R matrix;

\begin{listing}[!ht]
\begin{minted}{c}
  size_t MbyN = N * M;
#pragma omp target teams distribute parallel for
  for (size_t i = 0; i < MbyN; i++)
  {
    A_value[i] = scale * A_value[i];
  }
\end{minted}
\caption{Example of GPU-offloaded kernel using OpenMP directives.}
\label{list:ellpack_scale}
\end{listing}


In practice, OpenMP may not offer the ability to generate kernels with optimal performance on some devices, due to the inability to access certain fine-grained parallelism which is only available via vendor-specific methods, such as CUDA \cite{mohd2015gtc}. 
For this reason, we generally utilize a mixture of OpenMP, for data movement and execution of simple or non-performance-critical compute, with vendor-specific libraries (rocSparse, oneAPI, etc) for performance-critical kernels (see Sections \ref{se:amd}, \ref{se:mkl} and \ref{se:hypre}). 
Particularly for sparse matrices, we find that native OpenMP is unable to match the performance of vendor-optimized functions (see Figure \ref{fig:rocsparse_timings}). 
Fortunately, it is relatively easy to pass data to vendor libraries through raw pointers in C as allocated by OpenMP on the GPU. In these cases, we utilize the OpenMP \verb|use_device_ptr| functionality to perform compute on the device-side data arrays previously allocated. For example, Listing \ref{list:mklssyev} shows the interface to call an eigensolver using Intel oneAPI MKL libraries; 

\begin{listing}[!ht]
\begin{minted}{c}
#pragma omp target enter data \
  map(alloc:work[0:lwork])
#pragma omp target variant dispatch \
  use_device_ptr(evecs, evals, work)
  ssyev("V", "U", &N, evecs, &N, \
    evals, work, &lwork, &info);
\end{minted}
\caption{Code example illustrating a library call within an OpenMP region, in this case calling MKL eigensolver.}
\label{list:mklssyev}
\end{listing}



In some cases, we have utilized entirely separate workflows, such as MAGMA, which encompass both data movement and compute, bypassing OpenMP (see Section \ref{se:magma}). 

For sparse matrix functionality, most vendor-supplied libraries only use the CSR format, both on CPU and accelerator. In contrast, BML uses CSR (slightly modified), ELLPACK-R and ELLBLOCK formats. In order to leverage the performance of the vendor-supplied libraries, we therefore developed a set of functions to perform the appropriate data transformations. Here we show a simple example of transforming data from a canonical CSR format to ELLPACK-R on the GPU;

\begin{listing}[!htb]
\begin{minted}{c}
#pragma omp target teams distribute parallel for
  for (int i = 0; i < A_N; i++)
  {
    A_nnz[i] = csrRowPtr[i + 1] - csrRowPtr[i];
    for (int j = 0; j < A_nnz[i]; j++)
    {
      int idx = csrRowPtr[i] + j;
      A_value[ROWMAJOR(i, j, A_N, A_M)]
        = csrVal[idx];
      A_index[ROWMAJOR(i, j, A_N, A_M)]
        = csrColInd[idx];
    }
  }
\end{minted}
\caption{CSR to ELLPACK-R conversion using OpenMP target 
 directives. Here ``ROWMAJOR'' is a C macro that returns the 1-d index of an element $(i,j)$ for a matrix of size ${\rm A}\_{\rm M} \times {\rm A}\_{\rm N}$.}
\label{list:csr2ellpack}
\end{listing}

Although these data transformations inevitably introduce some overhead, the cost is outweighed by the potential penalty of using sub-optimal compute for performance-critical functions such as sparse matrix multiplication in the native ELLPACK-R format using OpenMP.

\section{Offloading dense linear algebra solvers}
\label{se:dense}

\subsection{MAGMA on Nvidia and AMD GPUs}
\label{se:magma}
The MAGMA library\cite{magma,magma2014} offers most of the functionalities we need for the dense matrix format.
It essentially implements the whole set of functions typically available in BLAS and Lapack for the GPU, and more. It currently supports Nvidia and AMD GPUs, and is expected to support Intel GPUs in the near future. For offloading the dense matrix format to GPU in BML, the choice was made to rely heavily, and mostly, on MAGMA, for memory allocation, data movement and linear algebra operations on Nvidia and AMD GPUs.

As an exception, we observed that the diagonalization function available in the cuSolver library ``cusolverDnDsyevd'' from Nvidia significantly outperforms the MAGMA functions and have thus added an interface to it. Calling a cuSOLVER function from within a MAGMA code turns out to be easy since both codes work with simple C pointers, and a data array allocated by MAGMA can be directly used by any cuSolver function.
The rocSolver library is the AMD equivalent of Nvidia's cuSolver and is easy to integrate with MAGMA as well. However we did not find that the diagonalization function in that library to be more performant than the MAGMA one.

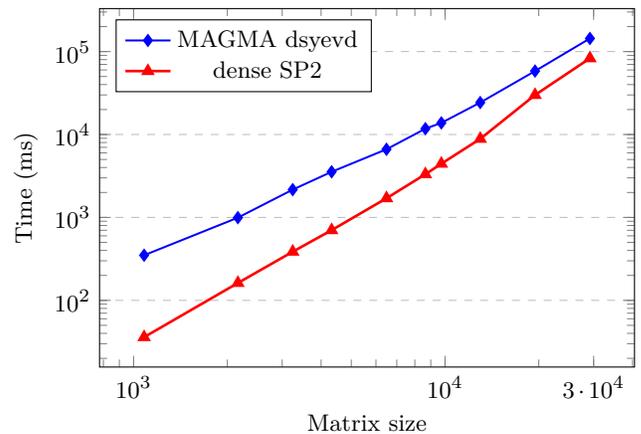
\begin{figure}
    \centering
    \begin{tikzpicture}
        \begin{loglogaxis}[width=.485\textwidth,height=2.5in,
                    legend pos=north west,
                    legend columns=1,
                    ylabel near ticks,
                    ylabel={Time (ms)},
                    xlabel={Matrix size},
                    ymajorgrids=true,
                    extra x ticks={30000},
                    max space between ticks=20,
                    grid style=dashed,
                    domain=1000:20000,
                    legend entries={ 
                    MAGMA dsyevd,dense SP2}]
        \addplot[blue,mark=diamond*, line width=0.75] table {data/crusher/magma_diag.txt};
        \addplot[red,mark=triangle*, line width=1.0] table {data/crusher/magma_sp2.txt};
        \end{loglogaxis}
    \end{tikzpicture}
    \caption{Comparison of time-to-solution between SP2 solver and dense diagonalization using MAGMA functionalities on AMD MI250X GPU for the PROGRESS 3D bio-benchmark.}
    \label{fig:magma}
\end{figure}

In Fig.~\ref{fig:magma}, we plot the time-to-solution for the SP2 solver (based on ``magma\_dgemm'' function) compared to dense diagonalization (using ``magma\_dsyevd\_gpu'' function) for the PROGRESS 3D bio-benchmark, as measured on AMD MI250X GPU.
SP2 significantly outperforms the dense diagonalization for all matrix sizes shown here (up to $N$=29,187), but more significantly for the smaller sizes.

Note that MAGMA provides a 2-stage eigensolver that could potentially be faster than the divide and conquer version we are currently using.
However its current interface only supports data on the CPU, and thus would require extra copies between the GPU and CPU in our implementation.

\subsection{MKL on Intel GPUs}
\label{se:mkl}

When offloading computation to Intel GPUs, we use the oneAPI MKL libraries \cite{oneapi}. At present we have only offloaded the computations for BML dense format, but we anticipate that the methodology will be directly transferable to the appropriate sparse formats. 
The offload of data to the Intel GPU is accomplished using OpenMP functionality, as described in Section \ref{se:openmp}. That is, we allocate the appropriate array, in this case a single dense block, on the GPU when the BML matrix is initialized. Subsequently, the correct state of the matrix is assumed to be that on the GPU, and all computation is performed on the GPU when possible in order to minimize data movement and to maximally leverage the improved GPU compute relative to CPU. 

The Intel model for GPU offload is enabled via offloaded versions of their oneAPI MKL libraries. The call signature is largely the same as the corresponding CPU function, with the addition of an appropriate OpenMP pragma. For example, the call for matrix addition on CPU (computing $\alpha A + \beta B$) in double precision is given by, using the BLAS C-interface:

\begin{listing}[!ht]
\begin{minted}{c}
    cblas_dscal(n, alpha, A->matrix, inc);
    cblas_daxpy(n, beta, B->matrix, inc,
                     A->matrix, inc);
\end{minted}
\caption{Computing $\alpha A + \beta B$ using the BLAS C-interface}
\label{list:cpu_mkl_blas}
\end{listing}

and the corresponding GPU offload is:

\begin{listing}[!ht]
\begin{minted}{c} 
double* A_matrix=A->matrix;
double* B_matrix=B->matrix;
#pragma omp target variant dispatch \
    use_device_ptr(A_matrix)
  cblas_dscal (n, alpha, A_matrix, inc);
#pragma omp target variant dispatch \
    use_device_ptr(A_matrix, B_matrix)
  cblas_daxpy (n, beta, B_matrix, inc, A_matrix, inc);
\end{minted}
\caption{Computing $\alpha A + \beta B$ using oneAPI MKL}
\label{list:gpu_mkl_blas}
\end{listing}

Note that for complex data types, the MKL CBLAS calls on the GPU require an ``\&'' in front of alpha and beta to get the address of a complex number.
Note also the \verb|target variant dispatch| and \verb|use_device_ptr| pragmas. Otherwise the call signature of the offloaded function appears identical to the corresponding host-side call. 

\begin{figure}[ht]
        \centering
        \begin{tikzpicture}
            \begin{loglogaxis}[width=.485\textwidth,height=2.5in,
                    legend style={at={(0.02,1.12)},anchor=north west},
                    legend columns=1,
                    ylabel near ticks,
                    ylabel={Time (ms)},
                    xlabel={Matrix size},
                    ymajorgrids=true,
                    extra x ticks={20000},
                    max space between ticks=20,
                    grid style=dashed,
                    domain=1000:20000,
                    legend entries={ 
                    GPU dsyevd,GPU SP2,
                    CPU 16 dsyevd, CPU 16 SP2,
                    CPU 52 dsyevd, CPU 52 SP2
                    }]
            \addplot[blue,mark=diamond*, line width=0.75] table {data/intel/1D_mkl_GPU_diag.txt};
            \addplot[red,mark=triangle*, line width=1.0] table {data/intel/1D_mkl_GPU_sp2_new.txt};
            \addplot[green,mark=diamond*, line width=0.75] table {data/intel/1D_mkl_CPU_16_diag.txt};
            \addplot[black,mark=triangle*, line width=1.0] table {data/intel/1D_mkl_CPU_16_sp2.txt};
            \addplot[yellow,mark=diamond*, line width=0.75] table {data/intel/1D_mkl_CPU_52_diag.txt};
            \addplot[cyan,mark=triangle*, line width=1.0] table {data/intel/1D_mkl_CPU_52_sp2.txt};
            \end{loglogaxis}
        \end{tikzpicture}
        \caption{PROGRESS benchmark timings measured on Intel Sunspot system for dense matrices of sizes N=1024 up to N=16384 using the oneAPI MKL library. Times are in milliseconds. CPU runs used either 16 or 52 threads, as indicated in legend. This work was done on a pre-production supercomputer with early versions of the Aurora software development kit.}
        \label{sunspot_timings}
\end{figure}
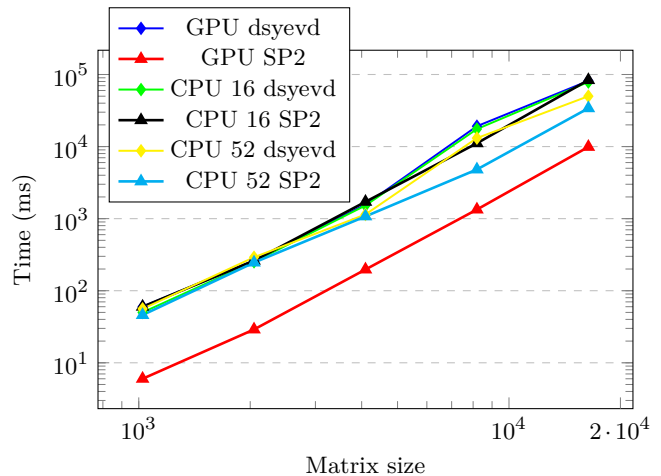

With the appropriate functions offloaded, we can then compare the performance of the SP2 algorithm on GPU with diagonalization, on both CPU and GPU. We used the PROGRESS soft matter synthetic Hamiltonians described in Ref.~\onlinecite{finkelstein2023fast} for this comparison, which is based on a simpler code implementation and allowed us to work around some of the compiler issues we faced on Intel GPUs available to us. The resulting timings on the Sunspot testbed are shown in Fig.~\ref{sunspot_timings}. Sunspot is a precursor to Aurora \cite{aurora_media}, where each node consists of 2 Intel Xeon CPU Max Series (codename Sapphire Rapids or SPR) and 6 Intel Data Center GPU Max Series (codename Ponte Vecchio or PVC). Each Xeon has 52 physical cores supporting 2 hardware threads per core. We utilize 16 threads per process on CPU, to simulate using $1/6$ of the CPU available per GPU. 
We also repeated the test with 52 CPU threads, equivalent to a full socket, which shows some improvement at larger sizes. 
As seen on other systems, the SP2 solver is performing well compared to the dense diagonalization, and GPU outperforms CPU. 

Although the choice to implement the offloaded functions with (almost) identical signatures to the CPU version makes the implementation straightforward from a coding standpoint, the lack of maturity of the software stack is an issue. 
Functions are being ported to the GPU gradually, and compiler updates may break working builds so frequent regression testing is needed. Overall, this is to be expected with a software stack in this stage of development. 

\section{Leveraging third-party libraries for $\mathcal{O}(N)$ sparse linear algebra solvers}
\label{se:sparse}

When it comes to iterative sparse solvers, such as SP2, the key kernel needed is a sparse-sparse matrix multiply. Compared to others such as sparse-dense matrix multiply, this function is not implemented in many libraries.
Previously, threaded sparse matrix methods in BML were implemented using the ELLPACK-R sparse matrix format. 
Our initial approach to GPU acceleration used OpenMP offload. Unfortunately this led to bottlenecks in matrix-matrix multiplications and additions -- the performance of the offloaded versions of these functions was lacking.
We thus switched to the use of third-party libraries for these numerical kernels.

\subsection{AMD rocSparse library}
\label{se:amd}

As faster methods were needed for AMD GPUs to prepare for the availability of Frontier at the Oak Ridge Leadership Facility\cite{Frontier}, we sought to replace these functions with AMD rocSPARSE methods. 
The CSR data used by rocSPARSE is obtained using an ELLPACK-R to CSR translation code within BML (see Section \ref{se:openmp}). Calls to rocSPARSE methods are then made within an $\verb|omp target data|$ region, to enable host or device pointers to be passed to rocSPARSE methods as appropriate. The code block in Listing~\ref{list:OffloadCode} illustrates the general approach of calling a rocSPARSE function using the GPU pointer to a BML data array (the approach is similar to the way the MKL methods are used on Intel GPUs as described in Section~\ref{se:mkl}). In this code block, the function $\verb|f()|$ performs a computation on $\verb|A_matrix|$ on the GPU. The $\verb|use_device_ptr(A_matrix)|$ clause instructs the compiler to pass the GPU pointer for $\verb|A_matrix|$ to $\verb|f()|$. 
The variable $\verb|A_N|$ is not included in the $\verb|use_device_ptr()|$ clause, indicating that the value on the host will be used. 

\begin{listing}[ht]
\begin{minted}{c}
double* A_matrix=A->matrix;
#pragma omp target data use_device_ptr(A_matrix)
{
  f(..., A_matrix,A_N,...)
}
\end{minted}
\caption{Using OpenMP offload for a rocSPARSE function ``f'' call and a BML matrix ``A''.}
\label{list:OffloadCode}
\end{listing}

The most computationally intensive kernel in SP2 is the sparse matrix multiplication; therefore, we first developed an interface to use the $\verb|rocsparse_spgemm()|$ function inside BML for this kernel. Several other bottlenecks subsequently were identified. In particular, the addition of two matrices in BML was accelerated using the function $\verb|rocsparse_csrgeam()|$. 

As of at least ROCm 5.3, the rocSPARSE methods require sorted column indices. This is important for our applications as the ELLPACK-R format is unsorted, and matrices can become unsorted during the course of calculations. For example, computing the matrix transpose leads to unsorted column indices. Moreover, even though the rocSPARSE methods require sorted matrices on input, they can produce unsorted matrices on output. We therefore changed the BML rocSPARSE code to sort matrices as needed. This was accomplished using the rocSPARSE $\verb|rocsparse_csrsort|$ function. In addition, to ensure that the resulting sparse matrices satisfy the thresholding criterion for including matrix elements, our implementation uses the rocSPARSE function $\verb|rocsparse_xprune_csr2csr|$). 

\begin{figure}
    \centering
    \begin{tikzpicture}
        \begin{loglogaxis}[width=.485\textwidth,height=2.5in,
                    ymin=10, ymax=200000,
                    legend pos=south east,
                    legend columns=1,
                    ylabel near ticks,
                    ylabel={Time (ms)},
                    xlabel={Matrix size},
                    ymajorgrids=true,
                    extra x ticks={30000},
                    max space between ticks=20,
                    grid style=dashed,
                    domain=1000:20000,
                    legend entries={OpenMP Offload,
                    MAGMA (dense),rocSPARSE}]
        \addplot[orange,mark=square*, line width=0.75] table {data/crusher/times_offload_3dbio_sort.txt};
        \addplot[blue,mark=diamond*, line width=0.75] table {data/crusher/magma_sp2.txt};
        \addplot[red,mark=triangle*, line width=1.0] table {data/crusher/timings_rocsparse_3dbio_sort_3500cols.txt};
        \end{loglogaxis}
    \end{tikzpicture}
    \caption{Comparison of time-to-solution of SP2 solver between sparse matrices using rocSPARSE and dense matrices using MAGMA on AMD MI250X GPU for the PROGRESS 3D bio-benchmark. OpenMP Offload timings for small matrix sizes are also shown, for comparison.}
    \label{fig:rocsparse_timings}
\end{figure}
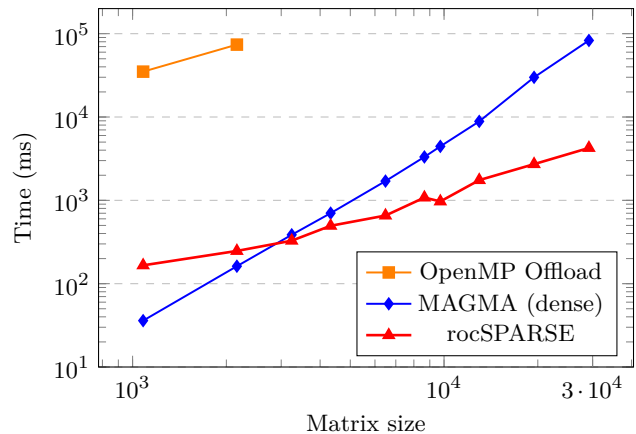

Figure~\ref{fig:rocsparse_timings} compares the performance of the SP2 solver for the PROGRESS 3D bio-benchmark using the rocSPARSE (red triangles) vs. MAGMA (blue diamonds) solvers. The builds used the Cray CCE compilers version 15, and AMD ROCm version 5.1. 
The rocSPARSE method shows approximate linear scaling, as expected for the sparse SP2 density matrix algorithms. Due to the approximate linear scaling, the rocSPARSE timings for large matrix sizes are smaller than the MAGMA timings (up to more than 10x). Two OpenMP Offload timings for small matrices are shown for comparison (Fig.~\ref{fig:rocsparse_timings}, orange squares). These data points show that the time required for the density matrix build using the rocSPARSE method is orders of magnitude smaller than that achieved for the OpenMP Offload code. Overall, the figure shows the substantial performance increase on GPU achieved when taking advantage of sparsity using rocSPARSE compared to the dense format using MAGMA for large matrix sizes. It also shows that rocSPARSE overcomes the performance limitations of the OpenMP Offload methods.

\subsection{hypre library} 
\label{se:hypre}

As mentioned above, functionalities for multiplying two sparse matrices by each other are not found in many linear algebra libraries.
The focus is often on solving sparse linear systems using iterative solvers, where multiplying a sparse matrix by a vector is the key ingredient.
That being said, electronic structure solvers are not the only ones using sparse-sparse matrix multiplications.
Another area where those are heavily used is in the algebraic multigrid community.
There, coarse grid operators are often computed as the product $A_L=R A_l P$, where $A_l$ is the discretized operator at the fine level, $A_L$ the discretized operator at the coarse level, and $R$ and $P$ are restriction and prolongation operators, which are used to coarsen or refine the data between levels.
All these operators are represented as sparse matrices, and thus the computation of $A_L$ is typically the result of two consecutive sparse-sparse matrix multiplications.

The hypre linear solver library \cite{hypre,FALGOUT2021} is a well-known open-source library with scalable Algebraic Multigrid solver capabilities, among other solvers. The sparse matrix-matrix multiplication routines in hypre provide access to both vendor library routines, as well as internal algorithms independent of the vendor. These internal algorithms are ported to HPC hardware with Nvidia, AMD and Intel GPUs based on their respective native programming paradigms. Thus, through the same interface to hypre, BML could access and utilize this performance critical sparse matrix functionality on different HPC platforms, leading to a performant and portable alternative to vendor-specific libraries. Additionally, leveraging open-source libraries such as hypre provides security through access to software capabilities and personnel with technical expertise and a shared interest in addressing software stack issues and challenges on emerging HPC hardware.

Accessing sparse matrix-matrix multiplication routines through hypre follows a similar approach to integrating BML with vendor libraries. First, matrix data is converted from ELLPACK-R format to standard CSR format on the device. hypre’s internal CSR matrix data structure uses raw C array pointers to store the matrix data. This interface makes it convenient to directly pass device pointers to standard CSR data, allocated by OpenMP, into hypre’s internal CSR matrix data structure without incurring additional overhead. Next, with the data on the device, the sparse matrix-matrix multiplication is performed using functionality provided by hypre. Finally, the result is converted back from CSR to ELLPACK-R using the same approach as for vendor libraries. Listing~\ref{list:hypre-matmat} provides a code snippet of how the integration with hypre is realized. 
Note that unlike rocSparse, hypre's interface does not put any requirement on the order of the elements in a CSR row.

\begin{listing}[!htb]
\begin{minted}{c}
    /* create hypre csr matrix */
    matA = hypre_CSRMatrixCreate( A_N,A_N,nnzA );
    matB = hypre_CSRMatrixCreate( B_N,B_N,nnzB );
#pragma omp target data use_device_ptr \
    (csrRowPtrA,csrColIndA,csrValA, \
    csrRowPtrB,csrColIndB,csrValB)
    {
       hypre_CSRMatrixI(matA) = csrRowPtrA;
       hypre_CSRMatrixJ(matA) = csrColIndA;
       hypre_CSRMatrixData(matA) = csrValA;

       hypre_CSRMatrixI(matB) = csrRowPtrB;
       hypre_CSRMatrixJ(matB) = csrColIndB;
       hypre_CSRMatrixData(matB) = csrValB;
    }
    /* perform matrix multiplication */
    matC  = hypre_CSRMatrixMultiplyDevice(matA, matB);
\end{minted}
\caption{Code snippet showing use of hypre for sparse matrix-matrix multiplication in BML. The assignment operations in the round brackets shows how data pointers are passed to hypre's internal data structure on device.}
\label{list:hypre-matmat}
\end{listing}

\begin{figure}
    \centering
    \begin{tikzpicture}
        \begin{loglogaxis}[width=.485\textwidth,height=2.5in,
                    ymin=100, ymax=20000,
                    legend style={at={(0.02,1.12)},anchor=north west},
                    legend columns=1,
                    ylabel near ticks,
                    ylabel={Time (ms)},
                    xlabel={Matrix size},
                    ymajorgrids=true,
                    extra x ticks={30000},
                    max space between ticks=20,
                    grid style=dashed,
                    domain=1000:20000,
                    legend entries={hypre (Nvidia V100), rocSPARSE (AMD MI250X), cuSPARSE (Nvidia V100)}]
        \addplot[red,mark=triangle*, line width=1.0] table {data/crusher/timings_hypre_nvidia.txt};        \addplot[blue,mark=diamond*, line width=0.75] table {data/crusher/timings_rocsparse_3dbio_sort_3500cols.txt};
        \addplot[orange,mark=square*, line width=1.0] table {data/crusher/timings_hypre_cusparse.txt};
        \end{loglogaxis}
    \end{tikzpicture}
    \caption{Time-to-solution of sparse $\mathcal{O}(N)$ SP2 solver using the hypre library on NVIDIA V100 GPU for the PROGRESS 3D bio-benchmark. For comparison, we also plot two data points when using the Nvidia cuSparse library on same V100 GPU, as well as the results using the rocSparse SP2 library on AMD MI250X GPU. 
    }
    \label{fig:hypre_timings}
\end{figure}
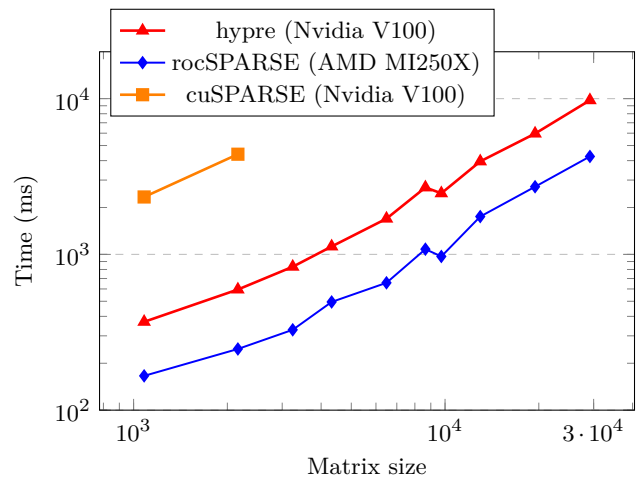

Initial challenges to this effort came from having a consistent compiler stack to enable an interoperable build of hypre with BML, with OpenMP Offload and vendor-specific compiler constraints. These challenges were eventually resolved.
Performance results for the PROGRESS SP2 benchmark using the sparse-sparse matrix multiplication from the hypre library are shown in Fig.~\ref{fig:hypre_timings} for an Nvidia V100 GPU.
Two data points using Nvidia cuSPARSE library are also shown. These correspond to the two smallest matrix sizes in the PROGRESS benchmark -- the only two we were able to complete using Nvidia CUDA 11 toolkit, due to the large memory requirements for the cuSPARSE sparse-sparse matrix multiplication.
They show a significantly better performance using hypre (factor 6-7X). 
The results from Section~\ref{se:amd} using rocSPARSE on AMD MI250X GPU, are also shown for reference. Taking into account the differences in hardware performance -- about 3X more flops and 1.8X better memory bandwidth for the AMD - GPU -- they indicate a comparable performance for hypre and rocSPARSE sparse-sparse matrix multiplications.

\section{Distributed memory solvers}
\label{se:dist}

Distributed memory approaches are very attractive for large problems that do not fit in the memory of a single node.
But they can also be used to speedup time-to-solution.
There is obviously always a cost to distributing a problem across multiple nodes due to communication and possibly extra computation for overlapping work.
But in many cases, these costs can be managed to a reasonable fraction of the compute time, and calculations can benefit from distributed resources.

Here we distinguish between two general ways of distributing an atomistic simulation: distributing the linear algebra problem of computing a DM, or dividing the physical system into a set of (potentially overlapping) subsystems.
The ScaLAPACK library\cite{blackford1997scalapack} is probably the most well-known library implementing distributed linear algebra operations.
In BML, we implemented a distributed matrix format which leverages the operations implemented for the various shared memory formats in BML.
This approach is described in Section \ref{se:distla}.
Partitioning the DM computation at the physical level was proposed by Yang and Lee in 1995 using a Divide and Conquer approach
\cite{YangLee1995}.
In Section \ref{se:graph}, we present a related approach implemented in PROGRESS.
It relies purely on matrix elements to determine the partitioning of the system, and is thus very appropriate for a library implementation totally independent of any specific electronic structure code.
It also leverages shared memory solvers implemented in BML, which are used for the linear algebra operations done for each physical sub-system.

\subsection{Distributed linear algebra}
\label{se:distla}

In order to leverage the implementation done for all the shared memory matrix formats in BML, we introduced a distributed matrix format where
each block owned by an MPI task is a BML matrix in a non-distributed (shared memory) format.
This allows us to have a distributed format for all the shared memory matrix formats already implemented in BML.
The restrictions we introduce with this format are that 
(i) the sub-matrices owned by each MPI task have to be square matrices,
(ii) the number of MPI tasks used has to be an integer squared.
We named that new format ``distributed2d''.
This format is built with the BML library when BML is configured to be built with MPI.

Our implementation is non-intrusive, leaving the shared memory matrix formats untouched.
It consists mainly of ``wrapper'' functions calling sub-matrix operations when possible, implementing the ``distributed2d'' matrix operations as combinations of “shared memory” matrix operations.
Some operations, such as the Frobenius norm for instance, need an MPI reduction at the end to get the global values.
Some operations (multiplication, transpose,…) require more substantial communications.
Our current implementation of distributed matrix-matrix multiplication is based on Cannon's algorithm.
Some operations are intrinsically more intrusive; for instance computing the bounds on the eigenvalues of a matrix using Gershgorin circles.
The strategy in this case is to add functionalities to the basic formats to be used by ``distributed2d''  implementation.
Some operations are beyond the scope of this project.
For instance, implementing a distributed eigensolver would require a lot of work beyond the resources of this project.
Fortunately, other libraries offer good distributed eigensolvers.
In BML, our eigensolver is thus simply interfacing with an existing solver.
For CPU, we have implemented an interface to ScaLAPACK\cite{blackford1997scalapack}.
For GPU, we have implemented an interface to the
ELPA library \cite{elpa2021}.

\subsection{Graph-based Divide electronic structure}
\label{se:graph}

Niklasson and collaborators recent findings\cite{Niklasson2016-gz} have demonstrated, both theoretically and practically, that there exists a bijective correspondence between matrix functions of sparse matrices and the same functions applied to only certain graph-restricted domains (parts) of the matrices. 
This theory enables the decomposition of the problem of computing a sparse matrix function into sub-problems involving much smaller dense matrices. The technique achieves near-perfect parallelism, where computations can be executed with distributed memory and minimal communication \cite{doi:10.1137/1.9781611974690.ch5, Ghale2017-xc}. The significance is twofold: it not only offers a systematic approach to addressing matrix function calculations by breaking them down into manageable components, but also capitalizes on the power of parallel computing to handle these computations concurrently. 

The original article used electronic structure calculations (the application of the Fermi-Dirac function) to validate this technique and was coined ``graph-based linear-scaling electronic structure theory.'' The utilization of electronic structure concepts to verify this theoretical result underscores its practical relevance in real-world applications, particularly in the domain of large-scale scientific computing.  Other researchers embraced this concept and translated it into libraries designed for the parallel application of matrix functions on massively large scales \cite{Dawson2018-oh}. 

Within the context of the PROGRESS library, we apply the Fermi-Dirac function to compute the system's DM. This technique however can be generalized to other functions. Given a Hamiltonian matrix $H$, and a threshold $\tau$, one can create a graph $G$ by constructing its adjacency matrix $A$ defined as

\begin{equation}   
A_{ij} = 
     \begin{cases}
       1 &\quad\text{if } |H_{ij}| \ge \tau \text{ and } i\neq j\\
       0 &\quad\text{otherwise.} 
     \end{cases}
\end{equation}

This graph can be partitioned into several components (or parts) using various methods referred to as graph partitioning techniques \cite{Adoni2020-nm}. 
We define a partition $\Pi$ of graph $G$ into $m$ parts, as a collection of $m$ sets of nodes of $G$, $\Pi=\left\{\pi_1,... \pi_i, .. \pi_m\right\}$.

The PROGRESS library provides several options for graph partitioning.
Perhaps the most straightforward one is to simply divide the matrix index list into relatively regular segments. 
PROGRESS also offers partitioning through the use of the METIS library, which implements various algorithms based on multiple constraint partitionings \cite{metis}. 
In addition, PROGRESS provides a partitioning algorithm with the objective of minimizing the total number of arithmetic operations if a $\mathcal{O} (N^3)$ complexity would dominate mathematical operations performed over each set of nodes. 
PROGRESS also implements several variations of graph partitioning inspired by the Kernighan-Lin method \cite{Kernighan1970-jo}. Finally, there is a method in which a METIS partitioning is refined using a simulated annealing technique to minimize the number of operations in a $\mathcal{O} (N^3)$ complexity algorithm. 

Within the context of electronic structure, two concepts immediately follow the idea of a partition: \emph{core} partition and \emph{core-halo} partition.
A partition is called a \emph{core} partition,  $\Pi_c= \left\{\pi^{c}_1,... \pi^{c}_i, .. \pi^{c}_m\right\}$, if every node in $G$ belongs to one and only one of the $\pi^c_k \in \Pi_c$.  If we extend each core component $\pi^c_k \in \Pi_c$ to also include the neighboring nodes of every node in $\pi^c$, this defines a \emph{core-halo} partition, $\Pi_{ch} = \left\{\pi^{ch}_1,... \pi^{ch}_i, .. \pi^{ch}_m\right\}$. Thus if a node $l$ belongs to $\pi^{c}_k$, then every directly connected neighboring node $o$ (where $A_{l,o} = 1$) belongs to $\pi^{ch}_k$. 

Given a partition $\Pi$ of $G$, for every element $\pi_k \in\Pi$, we can extract a submatrix $h^{k}_{\alpha \beta} := H_{i \in \pi_k, j \in \pi_k}$. There is then a one-to-one mapping between indices of $h$ and the indices of $H$, that is if $s_k$ is the size (number of nodes) of $\pi_k$, $\alpha \in [1,s_k] \leftrightarrow i \in \pi_k$.
BML implements the functionalities to extract a submatrix $h$ from a global matrix $H$, and to map data from a submatrix $h$ into a global matrix $H$.

To prevent discontinuities, our approach involves extracting submatrices from core-halo partitions, applying the necessary functions to these submatrices, and then subsequently constructing the final matrices. This is accomplished by first constructing a core-halo partition from a core partition. The extension is based on an ``auxiliary'' matrix that defines the connectivity of the orbitals. This auxiliary matrix can be the Hamiltonian itself, the overlap matrix, or a previously computed DM. 
A previously computed DM can be obtained from previous time steps (or a previous SCF step) during a geometry optimization and/or a QMD simulation.
From this extension we can extract the submatrix $h^{k_{ch}}$ for each part $k$ and apply a matrix function to it. Once the function has been evaluated for every part $k$ we can then extract the submatrices corresponding to the ``core'' partition and map them back to a full system density matrix $P$, where $P \approx f(H)$.

\begin{listing}[!ht]
\begin{minted}{fortran}
!given a Hamiltonian matrix h_bml and a matrix g_bml
!describing the graph, compute DM d_bml 
!tau: threshold for graph partition
subroutine prg_graphSolver(h_bml,g_bml,d_bml,tau,...)
  !Constructing the graph
  call prg_get_graph(g_bml,tau,...)
  !Partitioning the graph
  call prg_metisPartition(g_bml)

  !loop over local parts
  do i= localPartMin(myRank), localPartMax(myRank)
    !Extracting Hamiltonian sub-matrices
    call bml_matrix2submatrix(h_bml,s_h_bml(i),...)
    !Solving the sub-problem
    call prg_build_DM(s_h_bml(i), s_d_bml(i)...)
    !Add solution of sub-problem to DM
    call bml_submatrix2matrix(s_d_bml(i),d_bml,...)
  end do
  !Collect local and remote DM parts
  prg_collectMatrixFromParts(d_bml,...)
end program
\end{minted}
\caption{General density matrix graph-based distributed solver routine as implemented in PROGRESS.}
\label{api}
\end{listing}

In Listing \ref{api} we present a general algorithm that applies this graph-based approach to computing the density matrix. 
The inputs are the Hamiltonian (\verb|h_bml|) matrix, and the auxiliary matrix (\verb|g_bml|) that serves as a guess for the connectivity of the orbitals. 
In our PROGRESS implementation, all the matrices in the interface are sparse matrices. 
The construction of the graph is handled using the BML ELLPACK-R format given the sizes of the adjacency matrices that are typically involved in large systems that need memory distributed techniques. 
The auxiliary matrix is converted into a weighted adjacency matrix by taking the absolute values of every entry and a thresholding operation is applied to control the extent of the resulting graph. 
The resulting graph is then used in a graph partitioning algorithm to get the parts. 
The graph parts are then used to define the Hamiltonian sub-matrices (dense BML matrices) that are then solved for independently and concurrently using, for example, a regular diagonalization method (\verb|prg_build_DM()|). 
From the several DM sub-matrices obtained, each MPI task will reconstruct locally a ``partially filled'' full DM, before the full DM is assembled by summing up all contributions using an MPI reduction operation. 

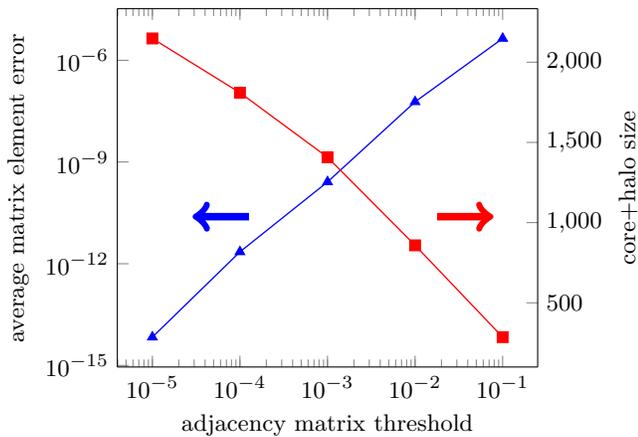
\begin{figure}
    \centering
    \begin{tikzpicture}
        \draw[line width=1mm,blue,<-] (1.,2.) -- (1.75,2.);  
        \draw[line width=1mm,red,->] (4.25,2.) -- (5.,2.);  
        \begin{loglogaxis}[width=.4\textwidth,height=2.5in,
                    ylabel near ticks,
                    ylabel={average matrix element error},
                    xlabel={adjacency matrix threshold},
                    axis y line*=left,
                    ]
        \addplot[blue,mark=triangle*,line width=0.5] table {data/graph_error.txt};
        \end{loglogaxis}
        
        \begin{semilogxaxis}[width=.4\textwidth,height=2.5in,
                    ylabel near ticks,
                    ylabel={core+halo size},
                    hide x axis,
                    axis y line*=right
                    ],
        \addplot[red,mark=square*,line width=0.5] table {data/corehalo.txt};
        \end{semilogxaxis}

    \end{tikzpicture}
    \caption{Error vs. thresholding parameter used to construct the adjacency matrix (blue) with average sizes of core+halo parts (red). These results were obtained from the PROGRESS benchmark Hamiltonian for $N$=2162 and a partition into 8 parts.} \label{fig:grapherror}
\end{figure}

    %
Fig.~\ref{fig:grapherror} shows the average DM element error and the average CH size as a function of the threshold value picked to build the adjacency matrix. We see that we get a well controlled error (Frobenius norm of the difference between the graph-based DM and the DM obtained using the dense diagonalization) by modifying the thresholding parameter used to construct the adjacency matrix. The average element error follows a linear function of the threshold on a log-log plot, indicating that the error is a polynomial of the threshold (error $\sim\tau^{2.2}$ in this case). When the submatrices are extracted, they contain a halo region (extra layer of surrounding orbitals) which is an extension from the cores (extracted set of atoms) arising from the overlapping graph-partitioning process. The smaller the threshold, the larger the overlap between the different parts and the smaller the error committed. On the other hand, the smaller the threshold, the larger the individual Hamiltonians that need to be solved independently and the higher the computational cost.

%

\begin{figure}
    \centering
    \begin{tikzpicture}
        \begin{loglogaxis}[width=.45\textwidth,height=2.5in,
                    log ticks with fixed point,
                    xmin=1000, xmax=5800,
                    legend pos=north west,
                    ylabel near ticks,
                    ylabel={wall-clock time (ms)},
                    xlabel={matrix size},
                    ymajorgrids=true,
                    grid style=dashed,
                    xtick=data,
                    legend entries={full diagonalization,graph-based decomposition}
                    ]
        \addplot[blue,mark=triangle*,line width=0.5] table {data/graph_scaling.txt};
        \addplot[red,mark=square*,line width=0.5] table {data/graph_scaling_gp.txt};
        \end{loglogaxis}  
    \end{tikzpicture}
    \caption{Wall-clock time vs. problem size for a fixed threshold of 0.01 for the graph-based solver using 8 parts (red). This is compared with the wall-clock time of a dense diagonalizaion of the entire problem (blue). Runs on a single Intel Core i7 CPU node with 12 OpenMP threads.} \label{fig:graphscaling}
\end{figure}
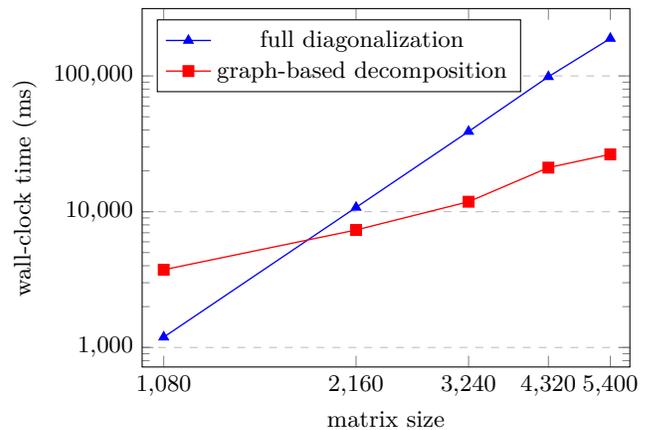

We tested this graph-based approach performance by evaluating the wall-clock time for the construction of DM as a function of the problem size.
In Fig.~\ref{fig:graphscaling}, we show the timings obtained and compared those to a standard dense diagonalization.
The computational cost as a function of problem size is remarkably promising. On a single Intel CPU node the graph-based approach has a very low prefactor scaling as compared to the regular DM construction method (see Fig.~\ref{fig:graphscaling}) and becomes competitive for matrix size $N$=2000 and beyond. 
We also see a larger speedup for larger system sizes.


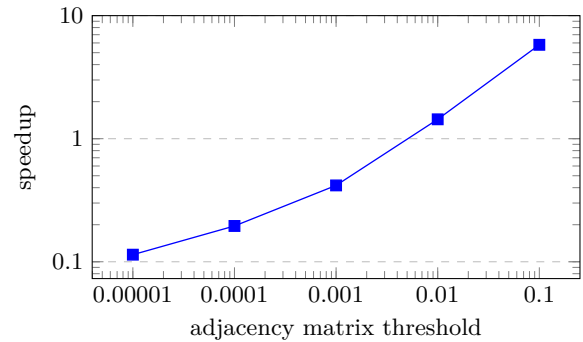
\begin{figure}
    \centering
    \begin{tikzpicture}
        \begin{loglogaxis}
       [width=.45\textwidth,height=2.in,
                    ymax=10.,
                    log ticks with fixed point,
                    legend pos=north west,
                    ylabel near ticks,
                    ylabel={speedup},
                    xlabel={adjacency matrix threshold},
                    ymajorgrids=true,
                    grid style=dashed
                    ]
        \addplot[blue,mark=square*,line width=0.5] table {data/graph_speedup.txt};
         \end{loglogaxis}
        
    \end{tikzpicture}
    \caption{Speedup with distributed graph solver compared to serial global solver as a function of the threshold used for graph partitioning. Results obtained for the PROGRESS benchmark with $N=2162$.} \label{fig:graphspeedup}
\end{figure}

With modern hybrid architectures, with GPU-accelerated nodes, speedups can be obtained only for larger problems.
The overhead associated with this distributed graph-based approach includes (i) partitioning the global matrix into ``core+halo'' parts, (ii) extracting the dense submatrices associated with each part from large sparse matrices, (iii) communications between MPI tasks to gather the calculated submatrices into the resulting global DM.
In Tab.\ref{tab:gp}, we show the matrix and submatrix sizes used by the graph-solver, as well as performance numbers obtained on Nvidia GPUs using a molecular system from the PROGRESS benchmark described in Section \ref{se:bml}. In this case, the difference between the time spent in the local solver and the total time for the distributed solver shows the time spent in the overhead operations (3.6 s). While it cannot be eliminated totally, we expect future code optimization to reduce it significantly.

\begin{table}[!htb]
\centering
\caption{\label{tab:gp} Performance of graph-based density matrix construction for a Hamiltonian matrix of size 12,972. 
Run on one node of Summit at OLCF (Nvidia V100 GPUs).
Local, single GPU solver is Nvidia cuSolver ``cusolverDnDsyevd''. The graph was constructed from the (precomputed) DM itself with a threshold of $10^{-2}$.
}
\begin{tabular}{c|c}
\hline
Number of parts &  4 \\ 
Number of MPI tasks &  4 \\ 
Number of GPUs &  4 \\ 
\hline
Matrix size & 12,972 \\
Number of core nodes/part & 3149--3314 \\
Number of core+halo nodes/part & 5657--5810 \\
\hline
Wall-clock time for local solver (s) & 2.0 \\
Total wall-clock time distributed solver (s) &  5.6  \\ 
\hline
Wall-clock time single GPU solver (s) & 8.7 \\
\hline
Speedup & 1.55 X \\ 
\hline
\end{tabular}
\end{table} 

\section{Concluding remarks}

Even though the software development environment using high-performance computing resources with GPU accelerators has improved substantially in recent years, it is still a challenge to produce software that is performant, portable and maintainable. 

With modern hybrid architectures where more than 90\% of the flops are those on the GPU, more and more scientific code needs to be developed and optimized for GPU execution. 
From a numerical point of view, maximal utilization of a GPU is both a lot of work and technically challenging, and may even require algorithm redesign. 
Dense eigenvalue solvers routinely used in the electronic structure community do not get the speedup one might expect on a GPU based on flops specifications, while other solvers based on matrix multiplications perform much better. 


In this paper, we demonstrated some ideas on how to address some of these issues, and described libraries (PROGRESS and BML) where these techniques are implemented.
From a performance point of view, when comparing the dense matrix-multiplication-based iterative solver SP2 with traditional dense diagonalization, we showed some significant speedups using AMD and Intel GPUs.
For $\mathcal{O}(N)$ solvers based on sparse SP2, focusing on Nvidia and AMD GPUs, we demonstrated how to leverage third-party libraries for core numerical kernels within an OpenMP offload implementation to achieve better performance than dense solvers for matrix sizes beyond $N=3,000$.
We also showed how some distributed memory solvers can be implemented, leveraging the performance of the shared memory operations implemented in BML. For these, the challenge remains to keep data transfer overhead low in comparison to on-GPU operations which are extremely fast on high-end devices.

From a practical point of view, the software stack can still be quite unstable and building a set of working libraries and code together remains a challenge.
In addition, OpenMP offload support differs from compiler to compiler which, with fewer debugging options than on CPUs, can require significant code development efforts.
While we expect software stack stability and reliability to improve with the maturity of the technology used in today's largest HPC resources, we also expect some challenges to persist for quite some time.

\section*{Acknowledgements}

This manuscript has been authored in part by UT-Battelle, LLC, under contract DE-AC05-00OR22725 with the US Department of Energy (DOE). The US government retains and the publisher, by accepting the article for publication, acknowledges that the US government retains a nonexclusive, paid-up, irrevocable, worldwide license to publish or reproduce the published form of this manuscript, or allow others to do so, for US government purposes. DOE will provide public access to these results of federally sponsored research in accordance with the DOE Public Access Plan (http://energy.gov/downloads/doe-public-access-plan).

This research used resources of the Oak Ridge Leadership Computing Facility at the Oak Ridge National Laboratory, which is supported by the Office of Science of the U.S. Department of Energy under Contract No. DE-AC05-00OR22725.
This research was supported by the Exascale Computing Project (17-SC-20-SC), a collaborative effort of the DOE Office of Science and the National Nuclear Security Administration (NNSA).

Portions of this work was performed under the auspices of the U.S. Department of Energy by Lawrence Livermore National Laboratory
under Contract DE-AC52-07NA27344

The authors would like to thank Rui Peng Li, LLNL, for his insightful comments and assistance in the integration of hypre into the BML library.

\section*{Author declarations}
The authors have no conflict of interest to disclose.

\section*{Data availability}
The data and code supporting the findings of this study are available from the corresponding authors upon reasonable request.

\bibliographystyle{unsrt}
\bibliography{references}
\end{document}